\documentclass[unsortedaddress,
twocolumn,showpacs,preprintnumbers,amsmath,amssymb]{revtex4}
\usepackage{graphicx}
\usepackage{dcolumn}
\usepackage{bm}
\usepackage{latexsym}
\begin{document}
\title{Traffic flow of interacting self-driven particles:\\
rails and trails, vehicles and vesicles}
\author{Debashish Chowdhury{\footnote{E-mail:debch@iitk.ac.in}}}
\address{Department of Physics, Indian Institute of Technology, Kanpur 208016, India.}

\vspace{0.7cm}

\begin{abstract}
One common feature of a vehicle, an ant and a kinesin motor is that they
all convert chemical energy, derived from fuel or food, into mechanical
energy required for their forward movement; such objects have been
modelled in recent years as {\it self-driven} ``particles''. Cytoskeletal
filaments, e.g., microtubules, form a ``rail'' network for intra-cellular
transport of vesicular cargo by molecular motors like, for example,
kinesins.  Similarly, ants move along trails while vehicles move along
lanes.  Therefore, the traffic of vehicles and organisms as well as that
of molecular motors can be modelled as systems of interacting self-driven
particles; these are of current interest in  non-equilibrium statistical
mechanics. In this paper we point out the common features of these model
systems and emphasize the crucial differences in their physical properties.
\end{abstract}

\maketitle

\section{\label{sec1}Introduction}

In the recent years non-equilibrium statistical mechanics has found 
unusual applications in research on traffic flow of various different 
types of objects. In this paper we consider mainly three different 
examples of such traffic, namely, (a) vehicular traffic 
\cite{css,helbing,naga}, (b) ant-traffic on ant-trails 
\cite{cgns,ncs,burd,couzin}, and (c) intra-cellular traffic of molecular 
motors carrying vesicular cargo moving along cytoskeletal filaments 
\cite{frey}.  Most of these models are essentially generalizations or 
extensions of the Asymmetric Simple Exclusion Process (ASEP) \cite{spohn}, 
which is, to our knowledge, the simplest model of systems consisting 
of interacting {\it driven} particles; the general aim of these 
investigations is to understand the interplay of self-organized 
{\it structures} and {\it transport} in systems driven far from 
equilibrium \cite{zia,schutz,stinchcombe}.

The aim of this article is (a) to summarize the main results of recent 
works on all the three systems mentioned above, elucidating the nature  
of various types of quenched randomness, (b) to present the challenging 
open problems, and (c) to indicate the possible trends of future 
developments in this frontier area of interdisciplinary research. 

The common modeling strategy is to represent the motile elements (i.e., 
vehicles, ants, molecular motors) by ``self-propelled'' particles which 
convert chemical energy (derived from fuel or food) into the mechanical 
energy required for the forward movement. In such generic models, the  
mutual influences of the motile elements on the movements of each other 
are captured by appropriate inter-particle interactions. In the spirit 
of the lattice gas models, the track for the traffic movement (i.e., 
the highway lane or ant trail or cytoskeletal filaments) are represented 
as discrete lattice of ``cells'' each of which can accomodate at most one 
partcle at a time. The dynamical laws governing the forward movement of 
the self-propelled particles in  such ``particle-hopping'' models are 
usually formulated as ``update rules'' in terms of cellular automata (CA) 
\cite{wolfram}. 

\section{\label{sec2}Vehicular traffic} 

The simplest model of interacting self-driven particles is the so-called 
totally asymmetric simple exclusion process (TASEP) \cite{spohn,zia,schutz}. 
Imposition of open boundary conditions leads to richer physics as 
compared to those for the corresponding model with periodic boundary 
conditions. The states of the system are updated either in parallel or 
in a random-sequential manner following rules which will be explained 
later in this section. 

\subsection{TASEP with periodic boundary conditions} 

\begin{figure}[tb]
\begin{center}
\includegraphics[width=0.8\columnwidth]{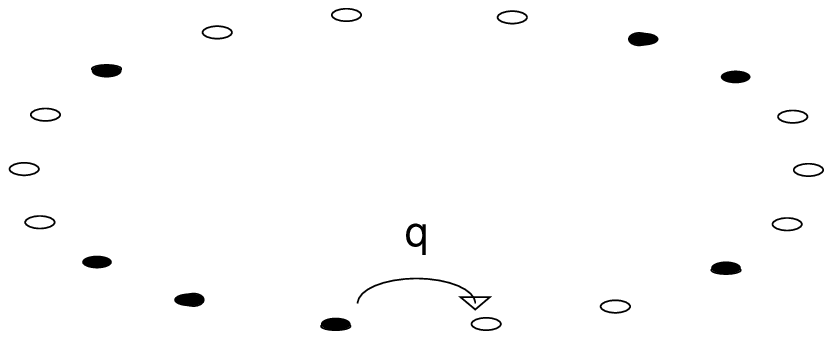}
\end{center}
\caption{TASEP with periodic boundary conditions. The hopping probability 
of the particles is $q$.}
\label{fig-1}
\end{figure}

\begin{figure}[tb]
\begin{center}
\includegraphics[angle=-90,width=0.8\columnwidth]{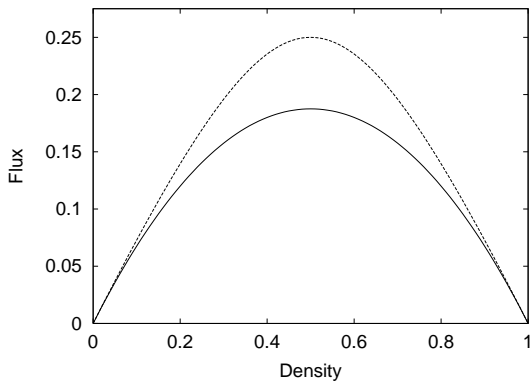}
\end{center}
\caption{Fundamental diagrams of the TASEP with periodic boundary 
conditions and (a) random-sequential updating (solid curve) 
(b) parallel updating (dashed curve), both for $q = 0.75$.}
\label{fig-2}
\end{figure}

\begin{figure}[tb]
\begin{center}
\includegraphics[width=0.8\columnwidth]{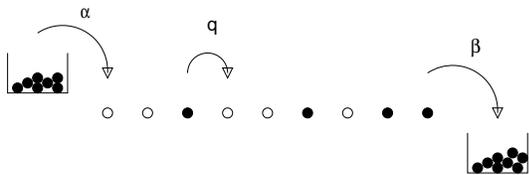}
\end{center}
\caption{TASEP with open boundary conditions.}
\label{fig-3}
\end{figure}

TASEP with periodic boundary conditions is sketched schematically in 
fig.\ref{fig-1}. 
In the original formulation of TASEP, the states of the system were 
updated in a random sequential manner where a particle is picked up 
randomly and moved forward by one lattice spacing, with the hopping 
probability $q$, provided the site in front of the particle is empty.

The Nagel-Schreckenberg (NS) model \cite{ns} is a minimal CA model 
of vehicular traffic on idealized single-lane highways; the maximum 
possible (discrete) speed of the vehicles is $V_{max}$. However, in 
the special case $V_{max} = 1$ this model reduces to the TASEP with 
parallel updating.

\begin{figure}[tb]
\begin{center}
\includegraphics[width=0.9\columnwidth]{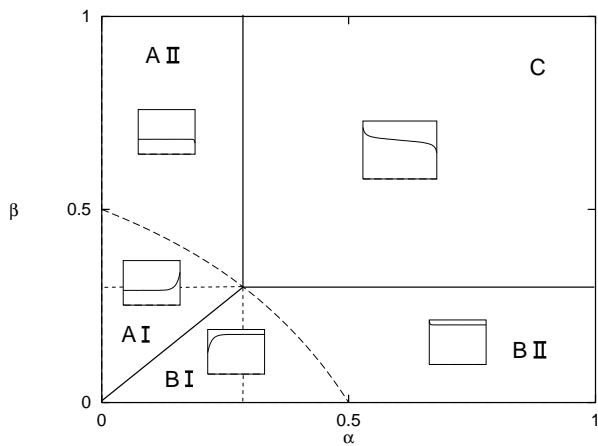}
\end{center}
\caption{ Phase diagram of the TASEP for {\it parallel} dynamics. 
The inserts show typical density profiles.  }
\label{fig-4}
\end{figure}

The exact fundamental diagrams of the TASEP with random-sequential 
updating and parallel updating are given by 
\begin{equation}
F_{r} = q ~ c ~(1-c) 
\label{eq-1}
\end{equation}
and
\begin{equation}
F_{p} = \frac{1}{2}\biggl[1 - \sqrt{1-4 q ~c ~(1-c)}\biggr], 
\label{eq-2}
\end{equation}
respectively; the two expressions (\ref{eq-1}) and (\ref{eq-2}), for 
identical hopping probability $q = 0.75$, are compared in 
fig.{\ref{fig-2}}. Note that both the expressions (\ref{eq-1}) and 
(\ref{eq-2}) exhibit particle-hole symmetry, i.e., these are symmetric 
under the interchange of $c$ and $1-c$. In section \ref{sec3}
we shall show how this symmetry is broken in a model of ant-traffic 
on ant-trails.

\subsection{TASEP with open boundary conditions} 

The open boundary condition is, however, closer to the real vehicular 
traffic on a stretch of highway.  If open boundary conditions are 
imposed on the TASEP, additional rules must be specified to regulate 
the entry and exit of the particles at the two boundaries of the 
finite system. Usually, these are specified as follows: if the first 
site at the open point of entry is empty it is filled with probability 
$\alpha$ whereas particles occupying the last site at the point of exit 
hop out of the system with probability $\beta$ (see fig.\ref{fig-3}).

The open boundaries break the translational invariance of the system 
and give rise to stationary states with non-trivial density profiles.
Such model systems have been investigated thoroughly over the last 
decade from the point of view of fundamental principles of 
non-equilibrium statistical mechanics.  In contrast to equilibrium 
systems with short-range interactions, such driven non-equilibrium 
systems can exhibit transitions from one dynamical phase to another, 
even in one-dimension with only short-range interactions, with the 
slight change of boundary conditions \cite{krug,css}. 

The typical phase diagrams of the TASEP with open boundary conditions 
are sketched in fig.\ref{fig-4}; the qualitative features of the 
phase diagram of TASEP is practically independent of the nature of 
the dynamics. In the low-density phase A the flux is independent of 
$\beta$ and limited only by $\alpha$. On the other hand, in the 
high-density phase B the flux is independent of $\alpha$ and determined 
by $\beta$. However, in the maximum flux phase C the current is 
independent of both $\alpha$ and $\beta$. Moreover, both the high- and 
low-density phases can be subdivided into two phases each, namely, 
AI, AII and BI, BII, respectively; these subphases are distinguished 
by the asymptotic behaviour of the density profiles at the boundaries 
(see the insets in fig.\ref{fig-4}). In section \ref{sec4} we shall see 
how this phase diagram is modified in models of intra-cellular transport 
of vesicular cargo by molecular motors.

\begin{figure}[tb]
\begin{center}
\includegraphics[width=0.8\columnwidth]{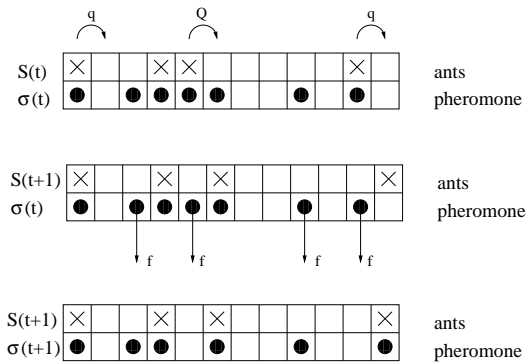}
\end{center}
\caption{ Schematic representation of typical configurations; it
also illustrates the update procedure. Top: Configuration at time $t$,
i.e.\ {\it before} stage $I$ of the update. The non-vanishing hopping
probabilities of the ants are also shown explicitly. Middle:
Configuration {\it after} one possible realisation of {\it stage $I$}.
Two ants have moved compared to the top part of the figure. Also
indicated are the pheromones that may evaporate in stage $II$ of the
update scheme.  Bottom: Configuration {\it after} one possible realization
of {\it stage $II$}. Two pheromones have evaporated and one pheromone has
been created due to the motion of an ant.  
}
\label{fig-5}
\end{figure}

\section{\label{sec3}Ant traffic on trails} 

Ants communicate with each other by dropping a chemical (generically 
called {\it pheromone}) on the substrate as they crawl forward 
\cite{wilson,camazine}. Although we cannot smell it, the trail
pheromone sticks to the substrate long enough for the other following
sniffing ants to pick up its smell and follow the trail. In our recent 
papers \cite{cgns,ncs} we have developed a particle-hopping model, 
formulated in terms of stochastic CA, which may be interpreted as a 
model of unidirectional ant-traffic on a trail. Rather than addressing 
the question of the emergence of the ant-trail, we have focussed 
attention on the traffic of ants on a trail which has already been formed.

Each site of our one-dimensional ant-trail model represents a cell
that can accomodate at most one ant at a time (see Fig.~\ref{fig-5}).
The lattice sites are labelled by the index $i$ ($i = 1,2,...,L$);
$L$ being the length of the lattice. We associate two binary variables
$S_i$ and $\sigma_i$ with each site $i$ where $S_i$ takes the value
$0$ or $1$ depending on whether the cell is empty or occupied by an ant.
Similarly, $\sigma_i =  1$ if the cell $i$ contains pheromone; otherwise,
$\sigma_i =  0$. Thus, in contrast to TASEP, we have two subsets of 
dynamical variables in this model, namely, $\{S(t)\}$ and $\{\sigma(t)\}$.
The instantaneous state (i.e., the configuration) of the system at
any time is specified completely by the set $(\{S\},\{\sigma\})$.

Since a unidirectional motion is assumed, ants do not move backward.
The forward-hopping probability of an ant is higher if it smells 
pheromone ahead of it. The state of the system is updated at each time 
step in {\it two stages}.  In our ant-trail model with {\it parallel} 
dynamics, at each stage the dynamical rules are applied {\it in parallel} 
to all ants and pheromones, respectively.\\

\noindent {\it Stage I: Motion of ants}\\[0.2cm]
\noindent An ant in cell $i$ that has an empty cell in front of it, i.e.,
$S_i(t)=1$ and $S_{i+1}(t)=0$, hops forward with
\begin{equation}
{\rm probability} = \left\{\begin{array}{lll}
            Q &\quad{\rm if\ }~\sigma_{i+1}(t) = 1,\\
            q &\quad{\rm if\ }~\sigma_{i+1}(t) = 0,
\end{array} \right.
\end{equation}
where, to be consistent with real ant-trails, we assume $ q < Q$.\\

\noindent {\it Stage II: Evaporation of pheromones}\\[0.2cm]
\noindent At each cell $i$ occupied by an ant after stage I
a pheromone will be created, i.e.,
\begin{equation}
\sigma_i(t+1) = 1\quad {\rm if\ }\quad S_i(t+1) = 1.
\end{equation}
On the other hand, any `free' pheromone at a site $i$ not occupied
by an ant will evaporate with the probability $f$
per unit time, i.e., if $S_i(t+1) = 0$, $\sigma_i(t) = 1$, then
\begin{equation}
\sigma_i(t+1) = \left\{\begin{array}{lll}
0 &\quad {\rm with\ probability\ } f,\\
1 &\quad {\rm with\ probability\ } 1-f.
\end{array} \right.
\end{equation}

We have also considered another version of our ant-trail model where 
the states of the system are updated in a random-sequential manner 
rather than in parallel \cite{ncs}. Note that in both the cases, 
because of the periodic boundary conditions, the dynamics conserves 
the number $N$ of ants, but not the number of pheromones.

This model is related to several other models. For example, in the 
limits $f \rightarrow 0$ and $\rightarrow \infty$ this ant-trail 
model reduces to TASEP with the hopping probabilities $q$ and $Q$, 
respectively. Moreover, the ant-trail model may be regarded as the 
opposite limit of the bus-route model \cite{loan,cd} (see ref.\cite{ncs} 
for the detailed comparison). Furthermore, the ant-trail model also 
has some similarities with the particle-hopping models of human trails 
of pedestrians \cite{helbped,burs}. 

\begin{figure}[tb]
\begin{center}
\includegraphics[width=0.8\columnwidth]{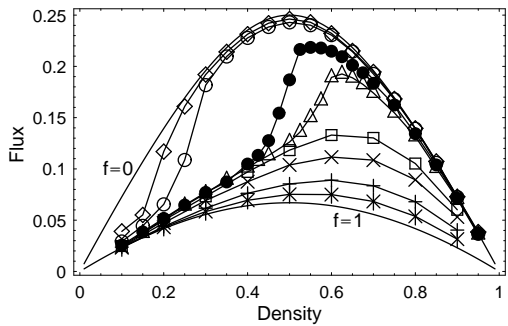}
\end{center}
\caption{Fundamental diagram of the ant-trail model with parallel 
updating for the parameters $Q = 0.75, ~q = 0.25$. The discrete 
data points corresponding to $f=0.0005 ({\Diamond})$, $0.001 (\circ$), 
$0.005 (\bullet)$, $0.01 ({\bigtriangleup})$, $0.05 ({\Box})$, 
$0.10 (\times)$, $0.25 (+)$, $0.50 (\ast)$ have been obtained from 
computer simulations; the lines connecting these data points merely 
serve as the guide to the eye..  }
\label{fig-6}
\end{figure}

\begin{figure}[tb]
\begin{center}
\includegraphics[width=0.8\columnwidth]{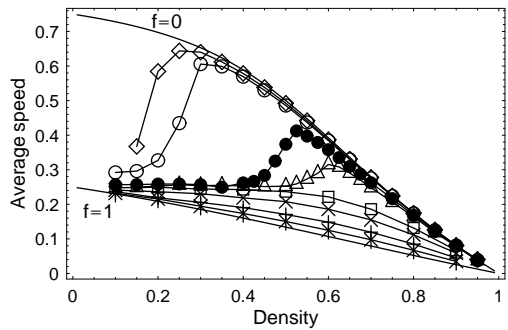}
\end{center}
\caption{Variation of the average speed of the ants in the ant-trail 
model with parallel updating.  Same symbols in figs.\ref{fig-6} and 
\ref{fig-7} correspond to the same values of the parameter $f$.}
\label{fig-7}
\end{figure}

The typical fundamental diagrams of the ant-trail model \cite{cgns} 
with parallel dynamics are shown in fig.\ref{fig-6}; the corresponding 
results in the case of random-sequential updating are qualitatively 
similar \cite{ncs}. The unusual shapes of the curves observed over a 
range of $f$ are consequences of the non-monotonic variation of the 
average speed of the ants with their density on the trail (see 
fig.\ref{fig-7}).

Both the ordinary mean-field theory (MFT), which accounts for the exact 
fundamental diagram of the TASEP with random-sequential updating, and 
$2$-cluster MFT \cite{ssni,css}, which succussfully predicted the exact 
fundamental diagram of the TASEP with parallel updating, fail to capture 
even the qualitative features of the fundamental diagrams of the 
ant-trail model shown in fig.\ref{fig-6} \cite{ncs}. However, a 
heuristic MFT, described in ref.\cite{ncs}, captures at least the 
qualitative features of the observed flow properties of our ant-trail 
model. 

\begin{figure}[tb]
\begin{center}
\includegraphics[width=0.8\columnwidth]{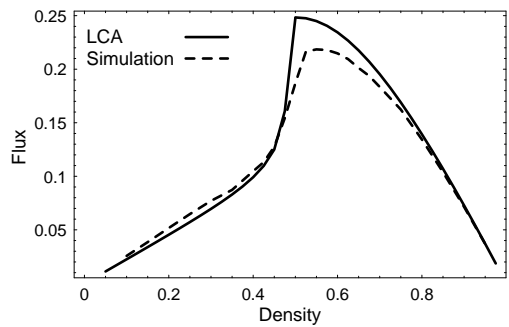}
\end{center}
\caption{Fundamental diagram ($f=0.005$) of the ant-trail model in 
the LCA (solid curve) s compared with the simulation data (broken 
curve).}
\label{fig-8}
\end{figure}

In order to develope a quantitative theory for the flow properties of 
the ant-trail model, we have analyzed the spatial organization of the 
ants by computer simulations. Analyzing these observations we concluded 
that in the anomalous regime, loose clusters of ants dominate; the term  
``loose'' means that there are small gaps in between successive ants in 
the cluster although the cluster appears to be an usual compact cluster 
if seen from a distance. As shown in fig.\ref{fig-8}, the fundamental 
diagram we calculated within the ``loose''-cluster approximation (LCA) is 
in good quantitative agreement with the corresponding data we obtained 
from computer simulations of the model.

\section{\label{sec4}Intra-cellular traffic of molecular motors} 

Molecular motors are protein molecules that convert the chemical 
energy, released by the hydrolysis of ATP, into mechanical energy  
required for its forward movement during intra-cellular transport 
of vesicular cargo \cite{howard}. The minimal models developed for 
explaining the mechanism of directed motion of isolated motor proteins 
are based on Brownian ratchets \cite{reimann}. In such models, each 
motor is represented by a particle. The essential features of the 
detailed mechano-chemistry of the molecular motor is captured in 
the Brownian ratchet models by a stochastic sequence of successive 
attachments and detachments of the motor with the cytoskeletal 
filamentary track (e,g, microtubule in the case of kinesin and dynein 
motors). In the simplest versions of these models \cite{julicher}, 
in the attached state, the particle representing a motor is subjected 
to a potential that is spatially periodic, but each period of which is 
asymmetric. In the detached state the particle executes an unbiased 
diffusive motion. In spite of its simplicity, such a minimal model can 
account for the directed, albeit noisy, movement of individual isolated 
motors.

To our knowledge, the question of the effects of interactions of the 
motors on the intra-cellular traffic was addressed theoretically for 
the first time only a few years ago \cite{menon}. In that work, the  
filamentary track was discretized in the spirit of the particle-hopping 
models described above and the motors were represented by field-driven 
particles. Both forward and backward movement of the particles were 
possible and the hopping probability of every particle was computed 
from the local potential. Thus, this model was a generalization of 
ASEP rather than TASEP where the hopping probabilities were obtained 
from the local potential which itself was time-dependent. The fundamental  
diagram of that model, computed imposing periodic boundary conditions, 
is very similar to those shown in fig.\ref{fig-2}. This observation 
indicates that further simplification in the model proposed in 
ref.\cite{menon} is possible to develope a minimal model for interacting 
molecular motors.  

\begin{figure}[tb]
\begin{center}
\includegraphics[width=0.8\columnwidth]{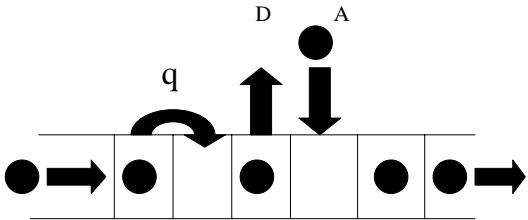}
\end{center}
\caption{Schematic representation of the model \cite{frey} of 
intra-cellular traffic of molecular motors carrying vesicular cargo.}
\label{fig-9}
\end{figure}

Recently, Parmeggiani et al.\cite{frey} have, indeed, developed such 
a minimal model for interacting molecular motors involved in 
intra-cellular transport by extending the TASEP with open boundary 
conditions. In this model, the molecular motors (e.g., kinesin or dynein) 
are represented by particles whereas the sites for the binding of the 
motors with the cytoskeletal tracks (e.g., microtubules) are represented 
by a one-dimensional discrete lattice.  Just as in TASEP, the motors are 
allowed to hop forward, with probability $q$, provided the site in 
front is empty. However, unlike TASEP, the particles can also get 
``attached'' to an empty lattice site, with probability $A$, 
and ``detached'' from an occupied site, with probability $D$ 
(see fig.\ref{fig-9}) from any site except the end points. The state 
of the system was updated in a random-sequential manner.

To my knowledge, this is the first application of TASEP to 
intra-cellular transport phenomena although it is not the first 
application of TASEP in the domain of biological systems; for example, 
a TASEP-like model was considered earlier for protein synthesis 
\cite{protein} 

\begin{figure}[tb]
\begin{center}
\includegraphics[width=0.8\columnwidth]{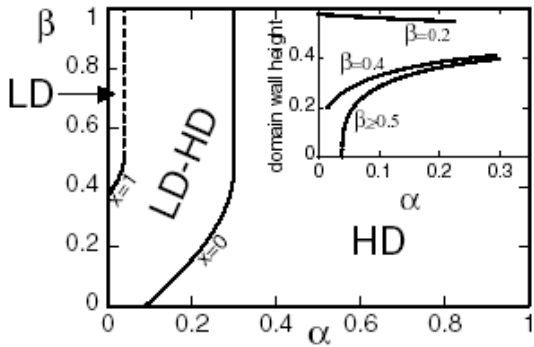}
\end{center}
\caption{The phase diagram of the model \cite{frey} of intra-cellular 
traffic of molecular motors carrying vesicular cargo (reproduced, with 
permission, from ref.\cite{frey}. The inset shows the dependence of 
the domain wall amplitude on $\alpha$ for different values of $\beta$.}
\label{fig-10}
\end{figure}

Carrying out Monte-Carlo simulations Parmeggiani et al.\cite{frey} 
demonstrated a novel phase where low and high density regimes, 
separated from each other by domain walls, coexist (see 
fig.\ref{fig-10}). Using a MFT, they interpreted this spatial 
organization as traffic jam of molecular motors.

\section{\label{sec5}Defects and disorder in particle-hopping models} 

\begin{figure}[tb]
\begin{center}
\includegraphics[width=0.8\columnwidth]{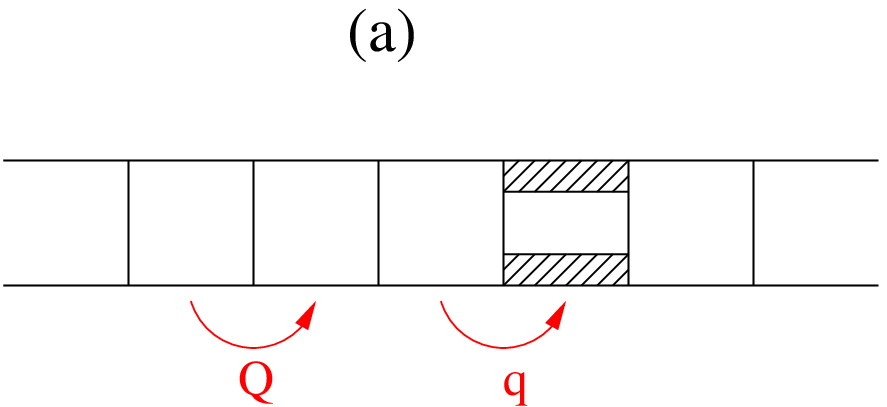}
\includegraphics[width=0.8\columnwidth]{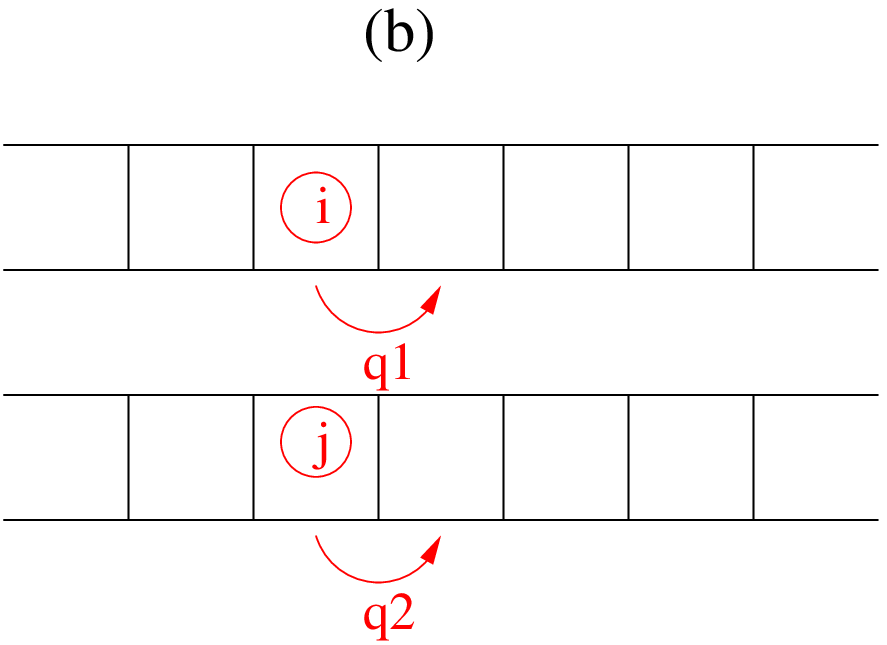}
\includegraphics[width=0.8\columnwidth]{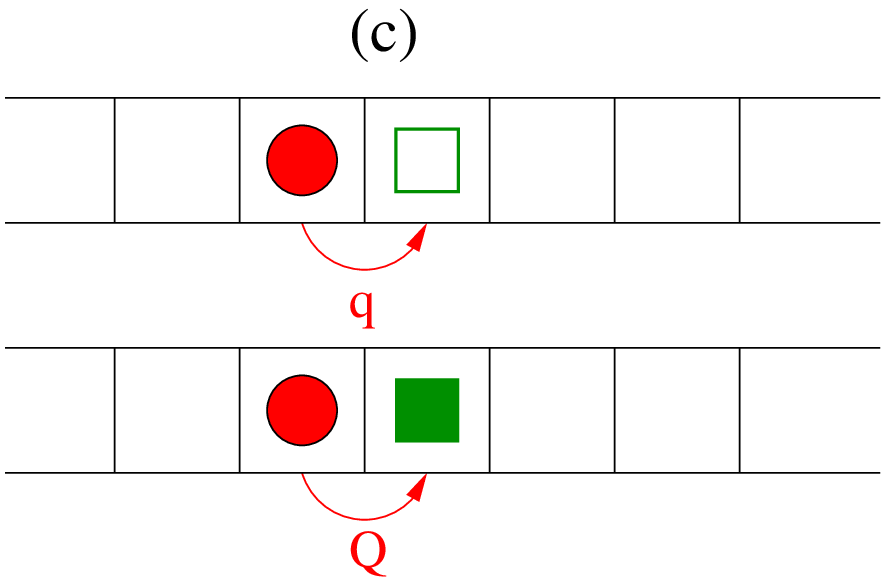}
\end{center}
\caption{Schematic representation of the different types of randomness 
in particle-hopping models. In (a) the randomness is associated with 
the track; the hopping probability $q$ at the bottleneck (partially 
hatched region) is smaller than the normal hopping probability $Q$, 
In (b) the randomness is associated with the particles; $q_1$ and $q_2$ 
being the time-independent hopping probabilities of the particles $i$ 
and $j$, respectively. In (c) the randomness arises from the coupling 
of the dynamics of the hopping particles (filled circle) with another 
non-conserved dynamical variable; the two possible states of the 
non-conserved variable are represented by open and filled squares.} 
\label{fig-11}
\end{figure}

At least three different types of defects and quenched randomness have 
been considered so far in the context of the models of interacting 
particles driven far from equilibrium. (a) First, the randomness may 
be associated with the {\it track} on which the particles move; 
typical examples are the bottlenecks on the roads (in the context of 
vehicular traffic) or defects on the microtubules (in intra-cellulat 
transport), etc. For example, as shown in fig.\ref{fig-11}(a), normal 
hopping probability at unblocked sites is $Q$ whereas that at the 
bottleneck is $q$ ($q < Q$). This type of quenched defect and disorder 
of the track leads to interesting phase-segregation phenomena (see 
\cite{css} for a review).

(b) The second type of randomness is associated with the hopping {\it 
particles}, rather than with the track. For example, the normal 
hopping probabilities of the particles may vary randomly from one 
particle to another (see fig.\ref{fig-11}(b)); the hopping probabilities 
are, however, ``quenched'' random variables, i.e., independent of time. 
In this case, the system is known to be exhibit coarsening of queues 
of the particles and the phenomenon has some formal similarities with 
Bose-Einstein condensation (reviewed in \cite{css}). 

Note that in case of the randomness of type (a), the hopping probability 
depends only on the spatial location on the track, independent of the 
identity of the hopping particle. On the other hand, in the case of 
randomness of type (b), the hopping probability depends on the hopping 
particle, irrespective of its spatial location on the track. In contrast 
to these two types of randomness, the randomness in the hopping 
probabilities of the particles in some models arises from the coupling 
of their dynamics with that of another non-conserved dynamical variable. 
For example, in the ant-trail model, the hopping probability of an  
ant depends on the presence or absence of pheromone in front of it (see 
fig.\ref{fig-11}(c)). Therefore, in such models with periodic boundary 
conditions, a given particle may hop from the same site, at different 
times, with different hopping probabilities. 

Defects of either (a) the cytoskeletal filaments or (b) the motor 
proteins or (c) the mechano-chemical coupling can cause malfunctiong 
of the intra-cellular transport leading to various types of diseases 
\cite{trafdis}. In order to get deep insight into the physical origin 
of such diseases, the recent model developed by Parmeggiani et al.
\cite{frey} has been extended \cite{chowetal}. This modeling strategy 
has opened up a new horizon for further unconventional applications 
of non-equilibrium statistical mechanics far beyond the traditional 
borderlines of physics.

\section{\label{sec6}Conclusion} 

In this paper we began with a brief introduction to TASEP \cite{zia,schutz} 
which is, perhaps, the simplest model of systems of interacting driven 
particles. TASEP, when updated in parallel, may be regarded as a special 
case, corresponding to the maximum allowed speed $V_{max} =1$, of the NS 
model \cite{ns}, the minimal model of vehicular traffic on single-lane 
highways. We have summarized some of the main known results on the 
{\it fundamental diagram} for the TASEP with periodic boundary conditions 
and the {\it phase diagram} of the TASEP with open boundary conditions. 
Then, we have shown how these results for TASEP get {\it qualitatively} 
modified by the generalizations or extensions required to model ant-traffic 
on ant-trails \cite{cgns,ncs} and molecular motor traffic on cytoskeletal 
filaments \cite{frey}. 

In the context of the ant-traffic on ant-trails, we have established how 
a combination of analytical and numerical calculations \cite{cgns,ncs} 
can account for the unusual shape of the fundamental diagram observed in 
the computer simulations of the ant-trail model. We have also presented 
the phase diagram obtained by Parmeggiani et al.\cite{frey} from studies  
of their recent model for molecular motor traffic. This phase diagram 
suggests the possibility of coexistence of high-density regions (traffic 
jam) and low-density regions (freely flowing traffic), separated from 
each other by domain walls, in a novel phase. Finally, we have mentioned 
some ongoing investigations on the effects of defects and disorder on 
molecular motor traffic \cite{chowetal}. This trend of research indicates 
the possibility of further unconventional, but very useful, applications 
of statistical physics in biological systems. 

\noindent {\bf Acknowledgements:} I dedicate this paper to Dietrich 
Stauffer on the occassion of his 60th birthday. During the two decades 
of our  collaboration he not only taught me the art of computer 
simulations but also inspired me to try unconventional applications of 
the conventional tools of statistical physics. I thank E. Frey, V. Guttal, 
F. J\"ulicher, A. Kunwar, K. Nishinari and A. Schadschneider for enjoyable 
recent collaborations and/or discussions. I also thank E. Frey for his kind 
permission to reproduce figure \ref{fig-10} from his paper \cite{frey}.
Part of this work was supported by the Alexander von Humboldt Foundation 
(under the re-invitation program for former fellows) and by the German 
Research Foundation (DFG) through a joint Indo-German research project.


\end{document}